\documentclass[jkps,final,fleqn,showpacs,showkeys,twocolumn]{revtex4-1}
\usepackage[pdftex]{graphicx}
\usepackage{amssymb}
\usepackage{amsmath}
\usepackage[hyperindex=true,pdftitle={High flux cold Rubidium atomic beam for strongly coupled Cavity QED},pdfauthor={Basudev Roy},colorlinks=true,
linkcolor=black,breaklinks=true,citecolor=blue,plainpages=false,
pdfpagelabels]{hyperref}
\usepackage{bm}
\usepackage{graphicx}
\usepackage{dcolumn}
\usepackage{float}
\usepackage{subfig}
\usepackage{wrapfig}
\usepackage{epstopdf}
\begin{document}
\setcounter{page}{0}
\title[]{High-flux Cold Rubidium Atomic Beam for Strongly-coupled Cavity QED}
\author{Basudev \surname{Roy}}\email{basudevroy@iiserkol.ac.in}
\affiliation{Indian Institute of Science Education and Research, Kolkata, India}\affiliation{Joint Quantum Institute, Department of Physics, University of Maryland, College Park and National Institute of Standards and Technology, MD, USA}
\author{Michael \surname{Scholten}}
\affiliation{Joint Quantum Institute, Department of Physics, University of Maryland, College Park and National Institute of Standards and Technology, MD, USA}

\date[]{Received 6 April 2007}

\begin{abstract}

This paper presents a setup capable of producing a high-flux continuous beam of cold rubidium atoms for cavity quantum electrodynamics experiments in the 
region of strong coupling.  A 2D$^+$ magneto-optical trap (MOT), loaded with rubidium getters in a dry-film-coated vapor cell, fed a secondary moving-molasses 
MOT (MM-MOT) at a rate  greater than 2 $\times$ $10^{10}$ atoms/s.  The MM-MOT provided a continuous beam with a tunable velocity. This beam was then directed 
through the waist of a cavity with a length of 280 $\mu$m, resulting in a vacuum Rabi splitting of more than $\pm$ 10 MHz. The presence of a sufficient number of atoms 
in the cavity mode also enabled splitting in the polarization perpendicular to the input. The cavity was in the strong coupling region, 
with an atom-photon dipole coupling coefficient g of 7 MHz, a cavity mode decay rate $\kappa$ of 3 MHz, and a spontaneous emission decay rate $\gamma$ of 6 MHz. 

\end{abstract}



\maketitle

\section{Introduction}
Any experiment requiring long interaction times between individual atoms and photons benefits from using slow atoms. This increases the probability of an interaction and enhances the effects being studied. One of the ways in which this interaction is practically implemented is in cavity quantum electrodynamics (cQED) where the photon bounces many times between the mirrors before escaping. Thus, the cavities provide a greater probability for the atom to interact with the photon. These systems are very useful in quantum information science, producing an entanglement between the incident photon and the atom present in the mode \cite{kimble0,mabuchi,zoller,rempe1}. These also enable study of quantum optics effects which are difficult to observe in free space \cite{carmichael0,orozco,orozco2}. Many realizations of cQED use slow-moving atoms \cite{kimble,rempe}.

One of the configurations for realizing interactions between a cavity and cold atoms is by dropping atoms cooled in a magneto-optical trap (MOT) into the cavity. This can be called the pulsed mode. In this mode, the MOT requires at least 1 second to capture sufficient atoms. These atoms last for only about a millisecond upon being dropped before they escape the cavity. Hence, the duty cycle for this process is very low. A continuous beam, on the other hand, can keep a steady atom-number at all times and provide a steady state for studying true quantum fluctuation effects in cQED \cite{carmichael2}, not to mention faster data taking.  

The ensemble trapped in a MOT has a typical diameter of 1 mm. A typical cavity has a spacing between mirrors of the order of 200 ${\rm \mu}$m. When the mode of the cavity has a waist (radius) of 30 ${\rm \mu}$m, the interaction area available to the dropped atoms is of the order of 0.2 mm $\times$ 0.06 mm = 0.012 ${\rm mm^2}$. Because the transverse cross-sectional area of the MOT is of the order of 1 ${\rm mm^2}$, geometrical considerations alone reduce the efficiency of the setup to 0.01.  Moreover, an atom moving at a speed of 5 m/s lasts for 10 ${\rm \mu}$s inside a cavity with a 30-${\rm \mu}$m waist. Thus, if the motion of the atoms alone is to be accounted and one effective atom is to be attained in the continuous mode, about $10^5$ atoms are required every second. The eventual efficiency drops to $0.01$ $\times$ $10^{-5} = 10^{-7}$. There are further losses that lower the efficiency even more. According to Carmichael and Sanders
\cite{carmichael}, for every 10 atoms of the atomic beam inside a cavity, there is only 1 effective atom. This happens because the atoms do not pass through the antinode of the cavity mode but are distributed over the entire standing wave. In addition to the distribution effect, the slow-moving atom beam also spreads in the transverse direction. Thus, accounting for the loss of another three orders of magnitude, the source needs to provide $10^{10}$ atoms/s to attain 1 effective atom. 

\begin{figure}[!h!t]
    \centering
{\includegraphics[height=70pt,width=250pt]{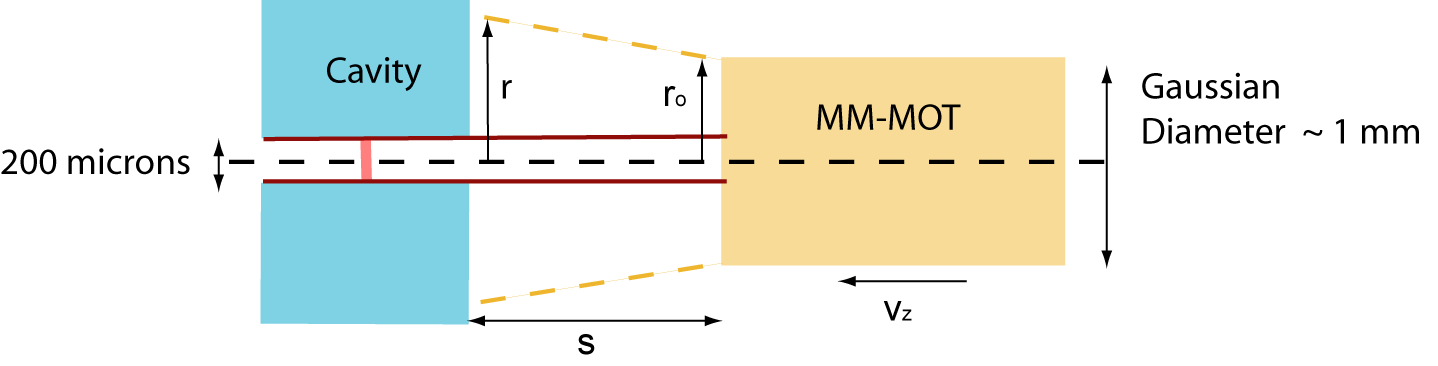}}\label{phfringes}
        \caption{Transfer of atoms from the MOT to the cavity. Atoms are released from a source of 1 mm diameter and have to pass through a cavity with a 200-$\mu$m opening and 30-$\mu$m waist. The geometry of the problem lowers the efficiency of the process to $\rm 10^{-2}$.}
 \label{geometry}
   \end{figure}

A static MOT with a push beam can only provide $10^8$ atoms/s \cite{pushbeam} and is incapable of being used as the source of cold atoms for strongly coupled cQED. Focusing techniques \cite{focusatoms} can narrow the transverse spread of the atomic beam, but  result in the loss of 1 to 2 orders of magnitude in the number of atoms during the process. Thus, those cannot be used to supply atoms in the continuous mode. A 2D$^+$ MOT \cite{prentiss} or a low-velocity intense source (LVIS) \cite{terra} have been proven to supply atoms at $10^{10}$ atoms/s and to generate a beam with a typical velocity of 15 m/s. If this velocity can be reduced to 2-3 m/s, the number of atoms in the cavity mode at all times can be significantly improved. Here, it may be reported that a cold continuous beam generated from an LVIS with secondary two-dimensional molasses has recently been used by Cimmarusti $et$ $al.$ \cite{andreas}. However, this two-dimensional molasses helps in improving the number of atoms present in the cavity at all times by only a factor of 2.
 
This paper presents a setup that generates a beam of cold rubidium atoms, recaptures and directs them through a strong coupling cavity with a speed of about 3 m/s. The paper is structured as follows: Section \ref{theory} gives a brief theoretical review. Section \ref{expt} mentions the experimental apparatus. Section \ref{res} describes the results of the experiment. Section \ref{data1} mentions the conclusions and discusses implications. 

\section{\label{theory}Theoretical review}

\subsubsection*{\label{MOTtheory}2.1 Moving Molasses MOT Theory}
The MOT uses a combination of a quadrupole magnetic field gradient and three pairs of opposing beams detuned from resonance to cool and simultaneously trap neutral atoms. The basic theory is mentioned in any standard textbook \cite{bill,metcalf}.

\begin{figure}[!h!t]
    \centering
{\includegraphics[height=180pt,width=180pt]{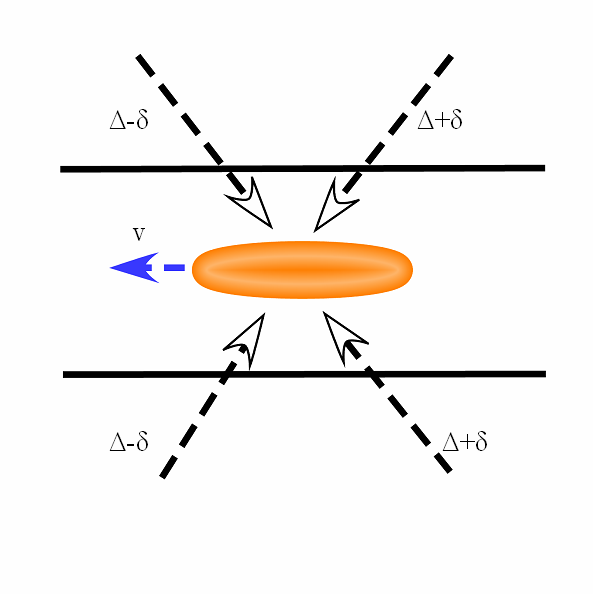}}\label{fft1}
        \caption{Beam detuning for a moving molasses MOT}
 \label{mmmot}
   \end{figure}

When the trapping is performed in a moving frame, the atoms move with the velocity with which this frame is moving with respect to the laboratory frame. Such a configuration is implemented by two of the trapping beams giving detuning ($\Delta$ - $\delta$), another pair giving detuning ($\Delta$ + $\delta$) and the third set giving $\Delta$, as shown in Fig. \ref{mmmot}. Here, $\Delta = {\omega-\omega_a}$, $\omega_a$ is the resonance frequency, and $\delta$ is the extra detuning. Such a configuration is called a moving molasses MOT (MM-MOT) \cite{thomann}. 

In a MM-MOT, the velocity of the atoms is given by 
\begin{equation}\label{eqn4}
2 \delta = {\bm k}\cdot{\bm v},
\end{equation}
where $\bm k$ is the wave vector of the cooling laser beam and $\bm v$ is the average velocity vector of the atoms. In the specific configuration of Fig. \ref{mmmot}, Eq. (\ref{eqn4}) becomes 

\begin{equation}\label{mmoteqn4}
 \lambda \delta /{\rm cos}(\pi/4)= v.
\end{equation}

Here, $v$ is the magnitude of the velocity vector ${\bm v}$.

\subsubsection*{\label{cavitytheory}2.2 Cavity Theory}

The cavity is a set of two mirrors which forms a Fabry-Perot interferometer between them. Every photon bounces many times between the mirrors before escaping. It builds up a mode where the electric field becomes sufficiently high to couple with the dipole moments of individual atoms. The coupling is given as 

\begin{equation}\label{eqn5}
 g = {\bm d}\cdot{\bm E_v}/\hbar, 
\end{equation}

where the field of a photon in a cavity with mode-volume ${\rm V_{eff}}$  is  ${\bm E_v} =\sqrt{\frac{\hbar c}{2 \epsilon_0 {\rm V_{eff}}}}$ and ${\bm d}$ is the dipole moment vector of the atom. The loss rates of the system are $\kappa$ and $\gamma$, and correspond to the light escaping through the mirrors and emitted via spontaneous emission, respectively. 

In the steady state, when $N$ maximally-coupled atoms interact with the mode of the cavity, the relation between the normalized field inside the cavity $y_1$ and the normalized input $x_1$ with the probe frequency $\omega$ is given by \cite{drummond}
\begin{equation}\label{eqn6}
y_1 = x_1 [ (1+ \frac{2C}{(1+{\Delta_1}^2+x_{1}^2)}) + i (\Theta - \frac{2C{\Delta_1}}{1+ {\Delta_1}^2+x_1^2})],
\end{equation}

where ${\Delta_1} = \frac{\omega-\omega_a}{\gamma/2}$, $\Theta = \frac{\omega-\omega_c}{\kappa}$, $\omega_c$ being the cavity resonance and $\omega_a$ the atomic resonance frequency. C is cooperativity given by $\frac{g^2}{\kappa\gamma}N$. 

In the presence of atoms inside the cavity, the levels of the system split (called vacuum Rabi splitting). The difference between the new split levels for the 1st excited state of the system is given by 
\begin{equation}\label{eqn7}
\omega_{VR} = 2 \sqrt{(g^2N- (\kappa-(\gamma/2))^2/4)}. 
\end{equation}

\begin{figure}[!h!t]
    \centering
{\includegraphics[height=250pt,width=250pt]{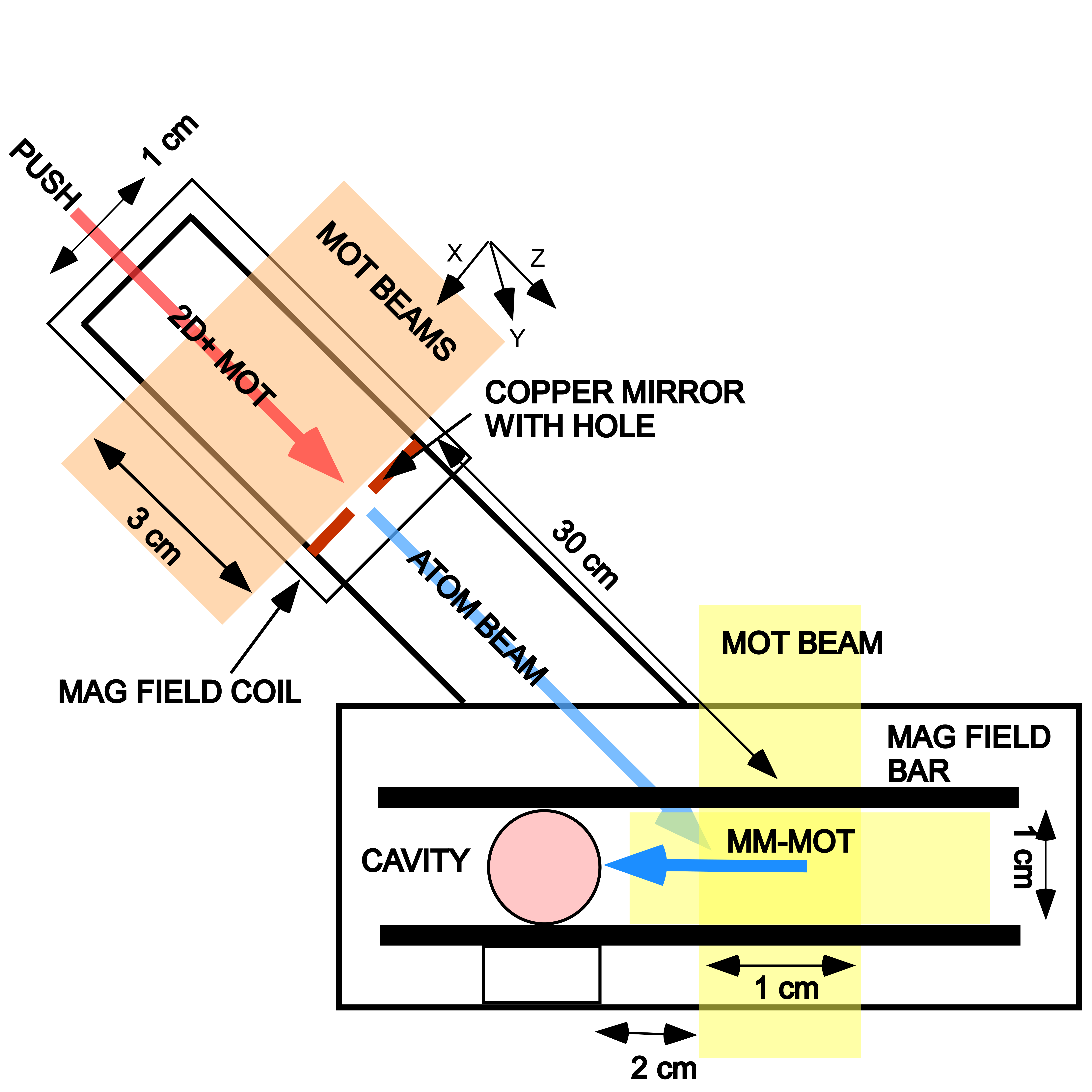}}\label{app}
        \caption{Experimental apparatus: The 2D$^+$ MOT is generated in a glass cell with one of the retroreflecting mirrors made from copper. The transverse MOT beams are elliptical, with a major axis of 3 cm and a minor axis of 1 cm. The axial beam is 1 cm in diameter with a hole diameter of 1 mm to match the size of the hole in the copper mirror. Atoms are guided through that hole by using an extra push beam. The MM-MOT has round-shaped beams 1 cm in diameter and sitting in a two-dimensional quadrupole magnetic field generated by bars with spacing of 1 cm. Atoms have to travel 2 cm from the end of the MM-MOT to the cavity. This distance was the closest attainable without getting the MOT beams clipped by the cavity mirrors.}
 \label{app}
   \end{figure}

\section{\label{expt}Experimental Apparatus}

Figure \ref{app} shows the experimental apparatus. It has mainly three parts, namely, the source, the MM-MOT and the cavity. The source of the atomic beam is a 2D$^+$ MOT. This configuration has a hole in one of the six MOT beams through which atoms are guided. It is implemented by making a hole in the mirror retro-reflecting one of the beams \cite{prentiss,unni}. The trapping beams along the X and the Y directions are elliptical with a length of 3 cm in the Z direction, but 1 cm in the X or the Y direction. The beam in the Z direction has a diameter of 1 cm with a hole of 1 mm. This is implemented by putting a mask of 1 mm to match the size of the hole in the copper mirror that reflects the Z-direction beam. An extra push beam is put through the hole to direct the atoms, as shown in Fig. \ref{app}. This configuration of the 2D$^+$ MOT is different from an LVIS because it employs an independent push beam but does not have a quarter-wave plate on the mirror in the Z direction. The MOT beams are obtained from a 780-nm diode laser (Toptica DLX110) having 500 mW with a line-width less than 1 MHz. The 780-nm laser light is locked to a 212-MHz transition below the target transition of $5S_{1/2} F=2$ to $5P_{3/2} F'$=3 in ${\rm ^{87}Rb}$ and is up-shifted using accousto optic modulators (AOM's) (Intraction, set at 97 MHz) in the double-pass mode to get a detuning of about 3 $\gamma$. The repumping laser addresses $5S_{1/2} F=1$ to $5P_{3/2}$ $F'$=2. The magnetic field gradient is about 12 Gauss/cm in the X and the Y directions. The rubidium is obtained from a dispenser wire placed inside the vacuum which releases atomic vapor as a current is passed through it. The intensity of each MOT beam is about 5 mW/${\rm cm^2}$. The walls of the glass cell used to generate the 2D$^+$ MOT are coated with a dry film layer that allow rubidium atoms to be thermalized on the surface and be reflected back into the MOT region. A silane-based mixture of dimethyldicholorosilane and methyltrichlorosilane, called SC-77, and an afterwash of methyltrimethoxysilane were used to generate the coating \cite{seth}. This helped in capturing more atoms for the beam at a lower rubidium vapor pressure.

The MM-MOT acts as a secondary MOT and recaptures the atoms in the beam. The individual frequencies required for the molasses trapping beams are also obtained from the 780 nm diode laser after being up-shifted using AOM's. Each of the beams in the MM-MOT are of 1-cm diameter. There is a two-dimensional magnetic field gradient of 14 Gauss/cm in the transverse directions of the MM-MOT, enabling the attainment of a cylindrical trap with a length of 1 cm and a diameter 1 mm when the detuning $\delta$ is made 0. When $\delta$ is made non-zero, this redirects the atoms towards the cavity at the velocity of the moving frame. The typical detuning $\Delta$ is 2 $\gamma$, with the extra detuning $\delta$ being varied from 2 to 5 MHz to generate velocities of 2 to 5 m/s. 

The velocity of the atoms in the 2D$^+$ MOT beam is typically 15 m/s.The flux of this MOT is inferred from the capture rate of the MM-MOT, with the detuning $\delta$ being made zero to trap in static frame. It has a typical value greater than $10^{10}$ atoms/s as can be seen from the initial loading rate of Fig. \ref{loadrate} showing the fluorescence of the MM-MOT. The initial loading rate can be called the flux of the 2D$^+$ MOT upon making the assumption that in the steady state operation of directing atoms into the cavity, the atoms are quickly released from the moving molasses after being recaptured. The MM-MOT uses a two-dimensional magnetic field gradient, thereby trapping in only those directions. For this reason, the MM-MOT requires four seconds to load, as seen in the 
Fig. \ref{loadrate}. 

\begin{figure}[!h!t]
    \centering
{\includegraphics[height=125pt,width=220pt]{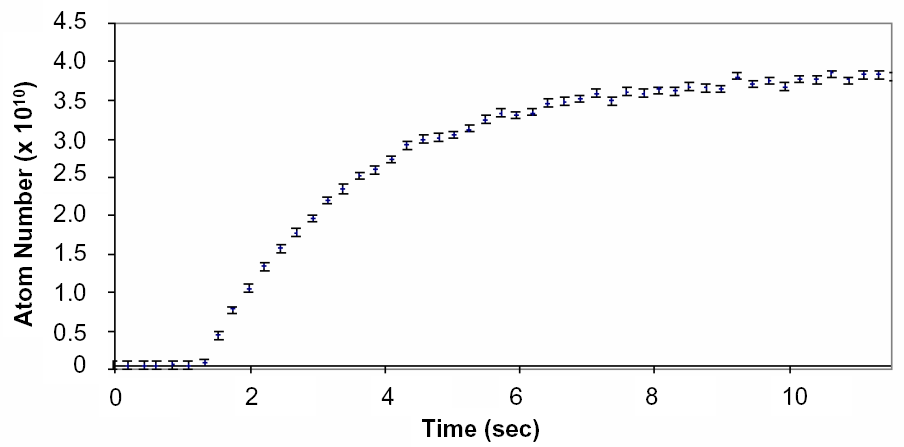}}\label{loadrate}
        \caption{Fluorescence from the MM-MOT after turning the 2D$^{+}$ atomic beam on when the detuning $\delta$ is made 0; the initial loading rate gives the flux of the 2D$^{+}$ beam}
 \label{loadrate}
   \end{figure}

The main cavity has two mirrors of 4-ppm transmission and a radius of curvature of 5 cm. The finesse of this cavity is 300000, the separation between mirrors is 280 $\mu$m and the saturation intensity is 1/8 photons. The probe is obtained from the 780-nm diode laser by using an electro-optic modulator (EOM) and selecting the upper sideband. (The laser is locked to a transition 212 MHz below the targetted transition. Hence, the upper sideband has to be used.) The intensity of light coming out of the cavity when there are no atoms present inside the mode (for 1/8 photons) is about 1 pW. Another cavity acts as an ultrastable reference for locking and is made up of a 50-cm mirror and a 5-cm mirror. It has a low finesse of about 15000 and a free spectral range of 3 GHz. This is made of Zerodur and has a very low coefficient of linear expansion with temperature. 

\begin{figure}[!h!t]
    \centering
{\includegraphics[height=250pt,width=260pt]{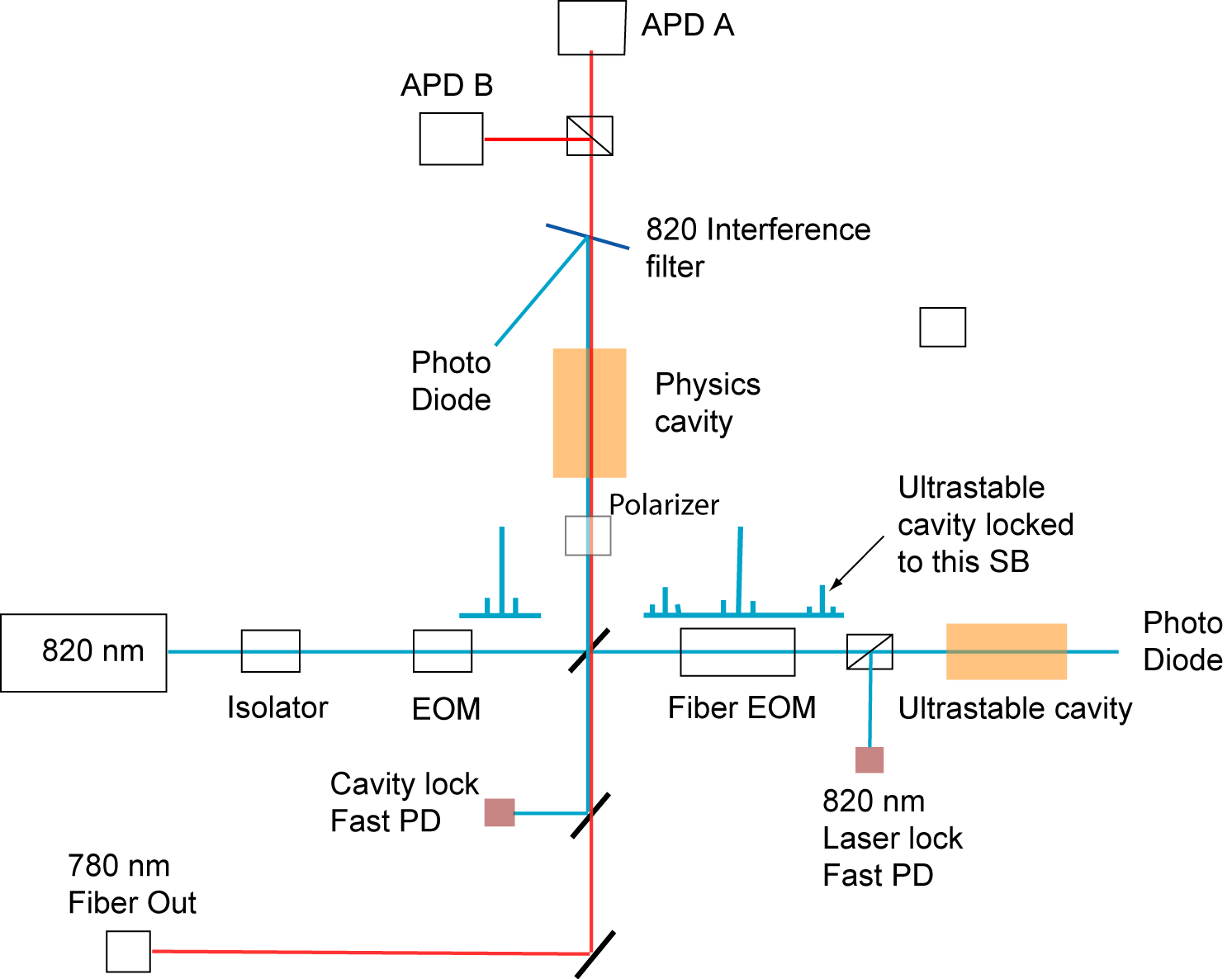}}\label{locking}
        \caption{Schematic diagram of the experiment along with the locking setup. The atoms are passed through the "physics cavity" while the ultrastable cavity is used for locking. An EOM is used to generate FM sidebands on the 820-nm laser light, which is then split, with one part being sent to the physics cavity. The other part of the 820-nm light is sent to an FM fiber modulator to generate sidebands tunable between 0 and 1.5 GHz which is then sent to the ultrastable cavity. The 820-nm laser is locked to this fiber modulator sideband at the ultrastable cavity, and the physics cavity is locked to the central peak of 820 nm. The fiber modulator sideband is then adjusted to allow the 780-nm light to pass through.}
 \label{locking}
   \end{figure}

We have used another diode laser at 820 nm for locking the cavity  (EOSI, giving about a 10-mW power and 1-MHz line-width). This 820-nm laser is first modulated using an EOM at 14 MHz and obtained FM sidebands. Part of this light is split and routed into the main cavity. The 820-nm light reflected from the cavity is then isolated and made incident on a fast photo-diode where it generates an error signal. The second part of the 820-nm light is passed through an FM fiber modulator (EOSpace, $V_{\pi}$ voltage about 2.5 V) and  modulated between 0 and 1.5 GHz  (half the free spectral range of the ultrastable cavity) to generate sidebands. This light is routed into the ultrastable reference cavity, and the reflection is utilized for generating the error signal. The sideband generated by the fiber modulator is locked to the ultrastable cavity, and the sideband frequency is changed to make the central peak of the 820-nm light copropagate with the 780-nm probe beam through the main cavity. The locking scheme is depicted in Fig. \ref{locking}.

The presence of the 2-D quadrupole magnetic field for the moving molasses MOT in the region of the cavity mode makes tracking the magnetic field axis difficult. The solution adopted is to add a bias  magnetic field along the direction of the moving molasses so that it does not change the alignment towards the cavity and yet provides a strong enough bias.

\section{\label{res} Results}

When the 2D$^+$ MOT source is used to feed the MM-MOT, a beam of atoms directed towards the cavity is obtained. In order to obtain the velocity of atoms coming out of the MM-MOT, we transferred the atoms from the lab frame to the moving frame suddenly by changing the detuning $\delta$ from zero to a finite value. This can be called the pulsed mode, as shown in Fig. \ref{pulsemode}. When $\delta$ is zero, the MM-MOT captures and traps the maximum number of atoms. Thus, the initial pulse of the mode has many more atoms as manifested by a dip in absorption as it goes through the cavity. The arrival time of the minimum of the dip provides information about the average velocity of the atoms. The typical distance from the end of the MM-MOT to the cavity is about 2 cm. This distance was the smallest attainable due to geometrical constraints of the system. The best signal, as seen in Fig. \ref{pulsemode}, is for a velocity of 3.2 m/s when in addition to the depth of the absorption dip, the width also becomes maximum. The atomic beam coming of the MM-MOT also drops under gravity. As the velocity of the atoms is reduced below 3 m/s, the effect of the drop becomes more prominent, resulting in fewer atoms arriving at the cavity mode. This results in less absorption, as seen in Fig. \ref{pulsemode}.  

\begin{figure}[!h!t]
    \centering
{\includegraphics[height=200pt,width=250pt]{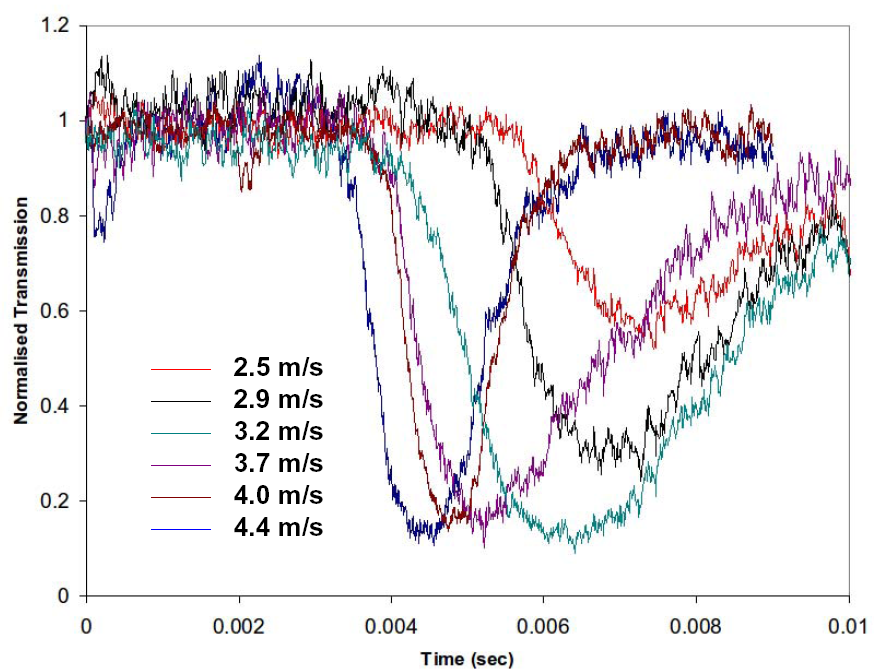}}\label{pulsemode}
        \caption{The detuning of the MM-MOT is changed suddenly from zero to a non-zero value to transfer the atoms from a static to a moving frame. This produces a pulse of atoms that is followed by the continuous beam. The time taken by the pulse to reach the cavity from the end of the MM-MOT (a distance of 2 cm) gives information about the velocity of the moving frame. For observation of the absorption in the pulsed mode, the dispenser current was reduced and kept constant at that value. The velocities mentioned are the calculated MM-MOT velocities.}
 \label{pulsemode}
   \end{figure}

The pulsed mode absorption signal is used to align the MM-MOT by moving the trapping beams to attain the largest dip. Once optimized, the dispenser current is gradually increased, and the molasses detuning is set to a constant value to get continuous beam. A linearly polarized probe laser beam is directed into the main cavity. The transmission of the probe in the same polarization as the input (called the driven mode), and the response in the orthogonal polarization is recorded (called the undriven mode), as seen in Fig. \ref{expt1}. 

\begin{figure}[!h!t]
    \centering
{\includegraphics[height=100pt,width=250pt]{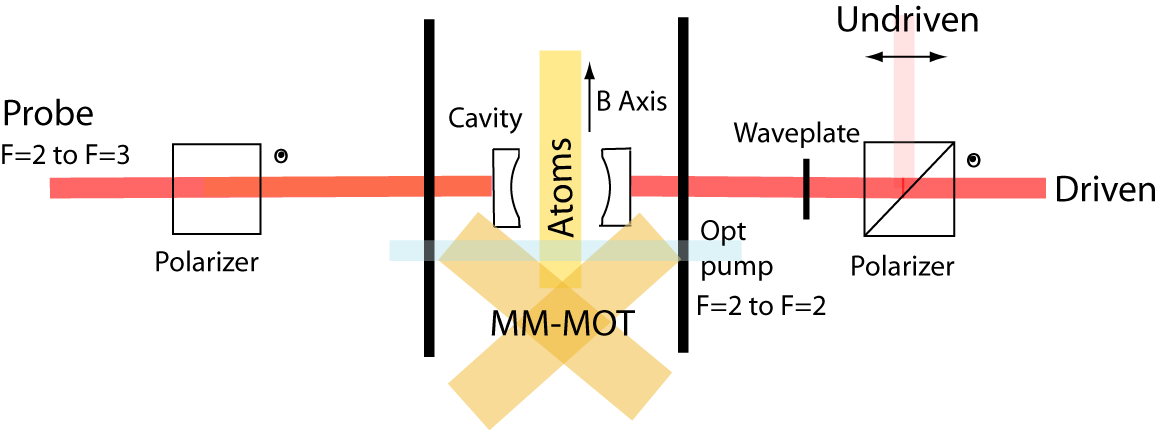}}\label{expt1}
        \caption{Atom beam directed into the cavity. The probe is addressed to the F=2 to F=3 levels in the D2 manifold of ${\rm ^{87}Rb}$. The driven mode is the response of the system to the input polarization and the undriven mode perpendicular to it.}
 \label{expt1}
   \end{figure}

In the experiment, a magnetic field bias is added in the direction of the atomic beam so that it does not affect the MM-MOT operation while enabling a well-defined magnetic field direction inside the cavity. The cavity detuning $\Theta$ is made zero by scanning the cavity lock frequency together with the probe laser frequency. Detection is performed using an avalanche photo diodes (APD). Equation (\ref{eqn6}) describes such a system when detection is performed in the driven mode. The values of the coupling coefficient $g$ (for linearly polarized light), the cavity decay rate $\kappa$ and the dipole spontaneous emission line width $\gamma$ are 7 MHz, 3 MHz and 6 MHz, respectively. The single-atom cooperativity is given as C = $\frac{g^2}{\kappa (\gamma/2)}$ = 5.4. As the dispenser current is increased, two Rabi sidebands appear gradually as shown in Fig. \ref{rabidata}. 

\begin{figure}[!h!t]
    \centering
{\includegraphics[height=150pt,width=225pt]{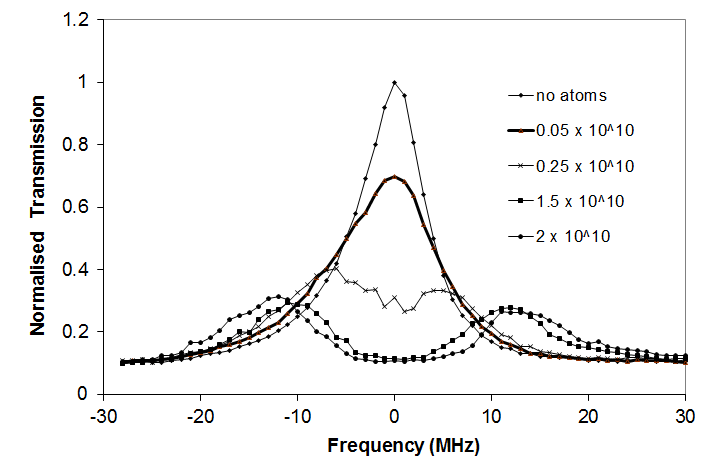}}\label{rabidata}
        \caption{Response of the system in the same polarization as the input (the driven mode). As expected from Vacuum Rabi splitting, the peak of the atom-cavity system splits into two as the number of atoms is increased upon increasing the flux of the atomic beam. The values mentioned are fluxes in atoms/s.}
 \label{rabidata}
   \end{figure}

The vacuum Rabi splitting for $N$ atoms, as given in Eq. (\ref{eqn7}), is $\omega_{VR}$ = 2 $g$ $\sqrt{N}$ because $\kappa$ = $\gamma/2$. The maximum splitting between the sidebands obtained experimentally, as shown in Fig. \ref{rabidata}, is 2 $\times$ 13 MHz. Thus, the number of effective atoms that contribute to the effect is $N$ = 3.4. 

\begin{figure}[!h!t]
    \centering
{\includegraphics[height=150pt,width=225pt]{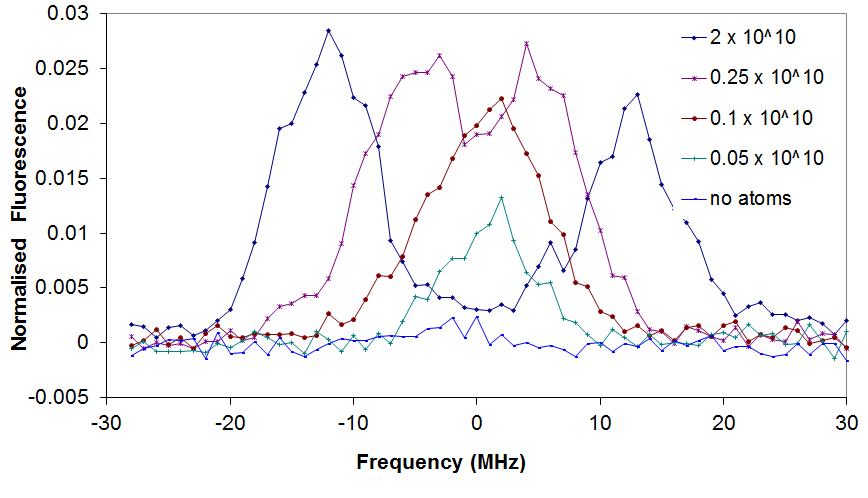}}\label{undriven}
        \caption{Response of the atom-cavity system for polarization orthogonal to the input (the undriven mode). The intensity is normalized to the height of the peak in the driven mode with no atoms in the cavity. The peak of the atom-cavity system in the undriven mode also splits into two as the number of atoms is increased. The values mentioned are the rubidium fluxes in atoms/s.}
 \label{undriven}
   \end{figure}

Figure \ref{undriven} shows the light in the undriven mode. The signal increases from zero and, with an increasing atom number, rises to about 0.025 of the peak of the empty-cavity driven mode before splitting into two. The light in the undriven mode is believed to be caused when the atoms get excited by linearly polarized light but emit circularly polarized photons while decaying to a different m-sublevel of the ground state, as mentioned in Ref. \cite{james}. The splitting in the undriven mode is the first such observation of the effect. Terraciano $et$ $al.$ \cite{terra} reported light in the undriven mode, but were unable to generate the high atom numbers required to see the splitting. The amount of fluorescence reported in that paper is also lower than that reported in the present work. The presence of a strong-coupling cavity may have influenced the high fraction of light in the undriven mode by collecting it and generating a mode for the atom to interact, thereby enhancing the effect \cite{james}. 

\section{\label{data1} Conclusions}
We have shown an apparatus to generate a high-flux continuous beam of atoms and to direct them into a strong-coupling cavity. The beam was strong enough to provide more than three effective atoms inside the cavity at all times. The setup could be improved further by directing the atom beam vertically upwards so that the velocity of the atoms going through the cavity would be reduced even further. This would allow more time for the atoms to interact with the mode. However, it would also make the cavity mounting scheme more complicated.

\section*{acknowledgements}
We thank the National Science foundation, the USA and the National Institute of Standards and Technology, the USA for their support. We also thank Luis Orozco for his help and support.  


\begin{references}


\bibitem{kimble0} Q. A. Turchette, C. J. Hood, W. Lange, H. Mabuchi and H. J. Kimble, Phys. Rev. Lett. {\bf 75}, 4710 (1995).
\bibitem{mabuchi} J. I. Cirac, P. Zoller, H. J. Kimble and H. Mabuchi, Phys. Rev. Lett. {\bf 78}, 3221 (1997).
\bibitem{zoller} K. M. Gheri, C. Saavedra, P. Torma, J. I. Cirac and P. Zoller, Phys. Rev. A {\bf 58}, R2627 (1998).
\bibitem{rempe1} T. Wilk, S. C. Webster, A. Kuhn and G. Rempe, Science {\bf 317}, 488 (2007).
\bibitem{carmichael0} D. G. Norris, L. A. Orozco, P. Barberis-Blostein and H. J. Carmichael, Phys. Rev. Lett. {\bf 105}, 123602 (2010).
\bibitem{orozco} D. G. Norris, E. J. Cahoon and L. A. Orozco, Phys. Rev. A {\bf 80}, 043830 (2009).
\bibitem{orozco2} M. L. Terraciano, R. Olson Knell, D. G. Norris, J. Jing, A. Fernandez and L. A. Orozco, Nat. Phys. {\bf 5}, 480 (2009).
\bibitem{kimble} A. Boca, R. Miller, K. M. Birnbaum, A. D. Boozer, J. McKeever and H. J. Kimble, Phys. Rev. Lett. {\bf 90}, 133602 (2003).
\bibitem{rempe} P. Munstermann, T. Fischer, P. Maunz, P. W. H. Pinske and G. Rempe, Opt. Comm. {\bf 159}, 63 (1999).
\bibitem{carmichael2} H. J. Carmichael, R. J. Brecha and P. R. Rice, Opt. Comm. {\bf 82}, 73 (1991).
\bibitem{carmichael} H. J. Carmichael and B. C. Sanders, Phys. Rev. A {\bf 60}, 2497 (1999).
\bibitem{pushbeam} L. Cacciapuoti, A. Castrillo, M. de Angelis and G. Tino, Eur. Phys. J. D {\bf 15}, 245 (2001).
\bibitem{focusatoms} P. N. Melentiev, P. A. Borisov, S. N. Rudnev, A. E. Afanasiev and V. I. Balykin, JETP Lett. {\bf 83}, 14 (2006).
\bibitem{prentiss} R. S. Conroy, Y. Xiao, M. Vengalattore, W. Rooijakkers and M. Prentiss, Opt. Comm. {\bf 226}, 20 (2003).
\bibitem{terra} M. L. Terraciano, R. Olsen Knell, D. L. Freimund, L. A. Orozco, J. P. Clemens and P. R. Rice, Opt. Lett. {\bf 32}, 982 (2007).
\bibitem{andreas} A. D. Cimmarusti, J. A. Crawford, D. G. Norris and L. A. Orozco, Rev. Mex. Fis. {\bf 57}, 29 (2011)
\bibitem{bill} S. Chu, W. Phillips and C. Cohen-Tannoudji, Rev. Mod. Phys {\bf 70}, 3 (1998).
\bibitem{metcalf} H. J. Metcalf and P. van der Straten, {\em Laser Cooling and Trapping}
 (Springer, New York, 1999).
\bibitem{thomann} P. Berthoud, E. Fretel and P. Thomann, Phys. Rev. A {\bf 60}, R4241 (1999).
\bibitem{drummond} P. D. Drummond, IEEE J. Quant. Elec. {\bf QE-17}, 301 (1981).
\bibitem{unni} S. Chaudhuri and S. Roy and C. S. Unnikrishnan, Phys. Rev. A {\bf 74}, 023406 (2006).
\bibitem{seth} S. Aubin, E. Gomez, L. A. Orozco and G. D. Sprouse, Rev. Sci. Intrum. {\bf 74}, 4342 (2003).
\bibitem{james} J. P. Clemens, Phys. Rev. A {\bf 81}, 063818 (2010).
\end{references}

\end{document}